\documentclass[ twocolumn]{aastex63}

\usepackage{graphicx}

\received{January 30, 2021}
\revised{June 10, 2021}
\accepted{December 31, 20XX}
\submitjournal{ApJ}

\shorttitle{Shock propagation in turbulent plasma}
\shortauthors{Carley et al.}
\graphicspath{{./}{figures/}}

\begin{document}

\title{Observations of shock propagation through turbulent plasma in the solar corona}

\correspondingauthor{Eoin P. Carley}
\email{eoin.carley@dias.ie}

\author{Eoin P. Carley}
\affiliation{Astronomy \& Astrophysics Section, Dublin Institute for Advanced Studies, Dublin 6, D02 XF86, Ireland.}

\author{Baptiste Cecconi}
\affiliation{LESIA, Observatoire de Paris, CNRS, PSL, Sorbonne Université, Meudon, France.}
\affiliation{Station de Radioastronomie de
Nançay, Observatoire de Paris, CNRS, PSL, Université d’Orléans, Nançay, France.}

\author{Hamish A. Reid}
\affiliation{Department of Space \& Climate Physics, University College London, UK.}

\author{Carine Briand}
\affiliation{LESIA, Observatoire de Paris, CNRS, PSL, Sorbonne Université, Meudon, France.}

\author{K. Sasikumar Raja}
\affiliation{LESIA, Observatoire de Paris, CNRS, PSL, Sorbonne Université, Meudon, France.}

\author{Sophie Masson}
\affiliation{LESIA, Observatoire de Paris, CNRS, PSL, Sorbonne Université, Meudon, France.}
\affiliation{Station de Radioastronomie de
Nançay, Observatoire de Paris, CNRS, PSL, Université d’Orléans, Nançay, France.}

\author{Vladimir Dorovskyy}
\affiliation{ Institute of Radio Astronomy, National Academy of Sciences of Ukraine, Kharkov,
Ukraine.}

\author{Caterina Tiburzi}
\affiliation{ASTRON, Netherlands Institute for Radio Astronomy, Postbus 2, 7990 AA, Dwingeloo, The Netherlands.}

\author{Nicole Vilmer}
\affiliation{LESIA, Observatoire de Paris, CNRS, PSL, Sorbonne Université, Meudon, France.}
\affiliation{Station de Radioastronomie de
Nançay, Observatoire de Paris, CNRS, PSL, Université d’Orléans, Nançay, France.}

\author{Pietro Zucca}
\affiliation{ASTRON, Netherlands Institute for Radio Astronomy, Postbus 2, 7990 AA, Dwingeloo, The Netherlands.}

\author{Philippe Zarka}
\affiliation{LESIA, Observatoire de Paris, CNRS, PSL, Sorbonne Université, Meudon, France.}

\author{Michel Tagger}
\affiliation{Laboratoire de Physique et Chimie de l'Environment et l'Espace, Université d'Orleans-CNRS, 45071 Orl\'{e}ans cedex 2, France.}

\author{Jean-Mathias Grie{\ss}meier}
\affiliation{Laboratoire de Physique et Chimie de l'Environment et l'Espace, Université d'Orleans-CNRS, 45071 Orl\'{e}ans cedex 2, France.}
\affiliation{Station de Radioastronomie de
Nançay, Observatoire de Paris, CNRS, PSL, Université d’Orléans, Nançay, France.}

\author{St\'ephane Corbel}
\affiliation{Station de Radioastronomie de
Nançay, Observatoire de Paris, CNRS, PSL, Université d’Orléans, Nançay, France.}
\affiliation{AIM, CEA, CNRS, Universit\'{e} de Paris, Universit\'{e} Paris-Saclay, F-91191 Gif-sur-Yvette, France}

\author{Gilles Theureau}
\affiliation{Laboratoire de Physique et Chimie de l'Environment et l'Espace, Université d'Orleans-CNRS, 45071 Orl\'{e}ans cedex 2, France.}
\affiliation{Station de Radioastronomie de
Nançay, Observatoire de Paris, CNRS, PSL, Université d’Orléans, Nançay, France.}
\affiliation{Laboratoire Univers et Th\'eories, Observatoire de Paris, Université PSL, CNRS, Université de Paris, 92190 Meudon, France}

\author{Alan Loh}
\affiliation{LESIA, Observatoire de Paris, CNRS, PSL, Sorbonne Université, Meudon, France.}

\author{Julien N. Girard}
\affiliation{AIM, CEA, CNRS, Universit\'{e} de Paris, Universit\'{e} Paris-Saclay, F-91191 Gif-sur-Yvette, France}

\begin{abstract}

Eruptive activity in the solar corona can often lead to the propagation of {shock waves}. In the radio domain
the primary signature of such shocks are type II radio bursts, observed in dynamic spectra as bands of emission
slowly drifting towards lower frequencies over time. These radio bursts can sometimes have inhomogeneous and fragmented
fine structure, but the cause of this fine structure is currently unclear. Here we observe {a type II
radio burst} on 2019-March-20th using the New Extension in Nan\c{c}ay Upgrading LOFAR (NenuFAR), a radio interferometer observing between 10-85\,MHz. We show  that the distribution of size-scales of density perturbations associated with the type II fine structure follows a power law with a {spectral index in the range of $\alpha=-1.7$ to -2.0}, which closely matches the value of $-5/3$ expected of fully developed turbulence. 
We determine this turbulence to be upstream of the shock, in background coronal plasma at a heliocentric distance of $\sim$2\,R$_{\odot}$. The observed inertial size-scales of the turbulent density inhomogeneities range from $\sim$62\,Mm to $\sim$209\,km. This shows that type II fine structure and fragmentation can be due to shock propagation through an inhomogeneous and turbulent coronal plasma, and we discuss the implications of this on electron acceleration in the coronal shock.

\end{abstract}


%
%

\section{Introduction} \label{sec:intro}

{Coronal mass ejections (CMEs) are eruptions of magnetized plasma from the solar corona into the heliosphere. These eruptions can drive shocks through the solar atmosphere}, and the primary radio signature of such shocks are known as type II radio bursts \citep{nelson1985, mann1996}. Type II bursts usually last tens of minutes and are characterised by bands of emission slowly drifting to lower frequencies over time. They can often show a fine structure which sometimes has the appearance of fragmented, short duration and narrow-band bursts of emission \citep{armatas2019}. It is expected that this fragmentation is likely due to the associated shock wave propagating through inhomogeneous coronal plasma \citep{afanasiev2009}, however the exact nature of the inhomogeneity has rarely been explored. A measure of the distribution of size-scales of the inhomogeneity may provide insight into the turbulent nature of shocks in the corona. Type II fragmentation may therefore be important in the study of coronal turbulence, as well as the implications of turbulence on particle acceleration in the coronal shock \citep{guo2010}.

Type II bursts are known to have a variety of different forms of sub-structure, which can come in the form of herringbones \citep{cairns1987, carley2013, carley2015}, as well as band-splitting of either fundamental or harmonic components of the radio burst \citep{vrsnak2001, chrys2018, maguire2020}. The bursts can also have a much less regular appearance, showing fragmentation and sporadic emission that can be broad or narrow band in frequency, particularly when they are observed with high time and frequency resolution dynamic spectra \citep{magdalenic2020}. Given that the corona and solar wind is known to be an inhomogeneous and turbulent medium \citep{bale2019, krupar2020}, the sporadic fragmentation of type II bursts may due to the turbulent nature of the medium through which the associated shock propagates. Some single event and statistical studies of the small scale structure of type IIs have been undertaken \citep{magdalenic2020, armatas2019}. However to our knowledge these properties have not been studied in the context of coronal turbulence. 

Theoretically, type II bursts are caused by plasma emission from beams of electrons accelerated at the shock front. The electrons are believed to be accelerated by the shock drift acceleration (SDA) mechanism \citep{holmon1983}, in which the electrons gain energy while undergoing a $\nabla B$-drift in the $\vec{v}\times \vec{B}$ convective electric field of the shock \citep{ball2001}. While SDA predicts particle energy gain upon single reflection from the shock, certain hybrid models employ a combination of SDA and turbulence to guarantee multiple reflections from the shock and hence higher energy gain \citep{burgess2006, guo2010}. This has been used to explain the $\sim$100\,keV energies of electrons observed at interplanetary shocks \citep{simnett2005}, which cannot be explained by a single-reflection SDA mechanism alone. Turbulent plasma and inhomogeneous shocks have also been suggested as an explanation for herringbone features in type II bursts \citep{zlobec1993, vandas2011}, which imply a time-variability or quasi-periodicity to the particle acceleration mechanism. However a complete explanation of herringbone time-variability still remains elusive.

{Turbulence and time variability of shock properties likely play an important role} in particle acceleration mechanisms in coronal shocks and the resulting appearance of type II sub-structure. It is only with modern radio instrumentation that we have the spectral resolution and sensitivity to probe coronal turbulence in type II fine structure as well as other burst types. {For example, \citet{chen2018} have recently used the Low Frequency Array \citep[LOFAR;][]{vanhaarlem2013} to show that type III burst fine structures have a power-law spectrum of intensity fluctuations with a spectral index of $\alpha=-1.71$. 
This suggests the radio burst fine structure could be related to the properties of the fully developed density turbulence through which the electron beam travels. } 
This mechanism was partly modelled using quasilinear theory of induced plasma emission in a turbulent coronal plasma \citep{reid2017}, {showing that Langmuir wave clumping in space can be directly related to background density turbulence}. {Using a numerical and analytical approach combined with LOFAR observations, \citet{reid2021} recently showed that type III fine structure properties are {induced} by coronal density turbulence and can be used as a remote probe of this turbulence.}

Here we employ the New Extension in Nan\c{c}ay Upgrading LOFAR \citep[NenuFAR;][]{zarka2012} to study the nature of coronal turbulence in the environment of {a complex type II burst during an eruptive event in the solar corona}. NenuFAR has unprecedented frequency resolution of 6\,kHz in its observing range of 10-85\,MHz. Given that the bandwidth of emission in frequency is related to the size scale of density structures in the corona, this allows us to study in detail the size distribution of density inhomogeneities responsible for fragmented type II emission. We show that this distribution specifically follows the signature of fully developed turbulence {during shock propagating through the solar corona}. In Section 2 we provide an observational overview of the type II burst, Section 3 we show the power spectral density analysis {for three different {parts} of the type II burst}, and we finally discuss the nature of the observed turbulence spectrum in Section 4.

\vspace{-0.1em}
%
%
\section{Observations}

 \begin{figure*}[!ht]
    \begin{center}
\includegraphics[scale=0.62, trim=3cm 17cm 0cm 0cm]{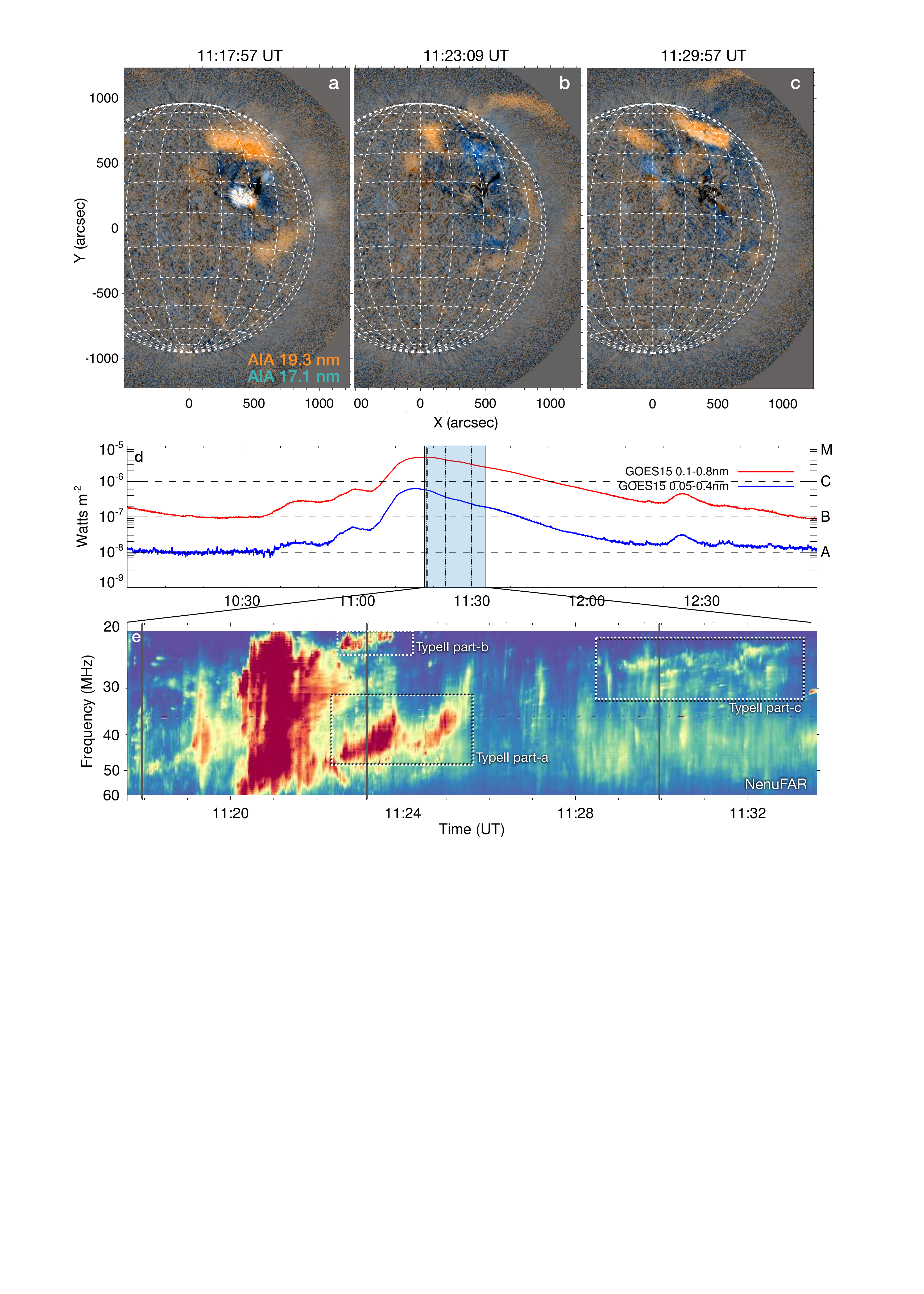}
\caption{(a,b,c) SDO 17.1\,nm (blue) amd 19.3\,nm (orange) running ratio observations of the eruption from AR12736. {An EUV wave is seen propagating both on-disk and off-limb, visible as enhanced emission in 19.3\,nm (orange regions)}. (d) GOES light curve of the C4.8 flare. The shaded region indicates the time interval of the dynamic spectrum and the vertical dashed lines are the times of the AIA images. (e) NenuFAR observations showing a summary of radio bursts taking place during the flare from 11:17:30--11:33:30\,UT. {The type II radio burst starts {in the NenuFAR frequency range } at 11:22\,UT and we have labelled three different {parts (type II parts-a,b,c)} which we examine separately}. {The solid vertical lines on the dynamic spectrum indicate the times of the AIA images.}}
\label{summary-fig}
\end{center}
\end{figure*}

On 2019-March-20th a C4.8 class flare took place in active region AR12736, peaking at $\sim$11:18\,UT, see Figure~\ref{summary-fig}. A faint EUV wave was observed during this time by the Atmospheric Imaging Assembly \citep[AIA;][]{lemen2012} 171\,\AA~ and 193\,\AA~ passbands, visible in Figure~\ref{summary-fig}a-c. To improve the wave visibility in the images we have used a combination of running ratio images (of 5 minute separation) and enhancement of low spatial frequency components using a Hanning window in the image Fourier domain. The wave is visible propagating both on disk and off-limb, where it propagates with a radial sky-plane speed of  {$480\pm150$\,km\,s$^{-1}$ and reaches the AIA field-of-view edge at $\sim$11:23\,UT. Deprojecting  this speed by the longitudinal angle of the source active region from the sky-plane (60$^\circ$), we find a speed of $950\pm310$\,km\,s$^{-1}$. The speed uncertainty results from a 50\,Mm positional uncertainty on the EUV wave}. The wave is then followed by the observation of a CME in the Large Angle Spectrometric Coronagraph \cite[LASCO;][]{brueckner1995} C2 field of view at 11:48\,UT\footnote{See \href{https://cdaw.gsfc.nasa.gov/CME_list/daily_movies/2019/03/20/}{https://cdaw.gsfc.nasa.gov/}}, which propagates at a constant speed of 500\,km\,s$^{-1}$ (in the sky-plane).

{The complex radio activity associated with this event started at $\sim$11:05\,UT with a patchy radio emission observed at $\sim$1000\,MHz\footnote{See \href{http://secchirh.obspm.fr/spip.php?page=survey&hour=1100&survey_type=1&dayofyear=20190320}{http://secchirh.obspm.fr}}.
This is followed by fast drifting type III bursts, a type II radio burst and a broad-band type IV continuum. The radio event was observed at metric to hectometric wavelengths, recorded by ground-based instruments such as the `Observations Radio pour FEDOME et l'\'Etude des \'Eruptions Solaires' (ORFEES) spectrometer and the Nan\c{c}ay Decametric Array \citep[NDA;][]{lecacheux2000}, as well as the space-based WIND/WAVES instrument \citep{bourgeret1995}. In this study we will exclusively focus on the NenuFAR observations.}


{From 11:18\,UT to 11:45\,UT} a series of complex radio bursts was observed by NenuFAR, {which provided a dyanmic spectrum from 20-55\,MHz during this period}. 
The radio bursts in NenuFAR begins with a number of type III bursts, a broadband feature starting at 11:20\,UT, followed by {a complex type II burst. In our analysis below, we examine different {parts} of the type II burst, labelled type-II part-a to part-c in the dynamic spectrum in Figure~\ref{summary-fig}e}. {Type II part-a} starts at 11:22\,UT at $\sim$45\,MHz and consists of two separate but connected series of herringbone bursts. {Type II part-b} is a small fragmented feature occurring at 11:22\,UT and starting at $\sim$23\,MHz (we show in the next section these are a fundamental-harmonic pair). 
{Type II part-c} is a faint and fragmented structure beginning at $\sim$11:28:30\,UT at $\sim$30\,MHz. Each part of the type II burst is {morphologically different. In} the following sections we analyse the size-scales of density inhomogeneity in the corona that were responsible for the features of each radio burst. 

%
%

\section{Methods \& Results}
Our goal is to attempt to identify any evidence of turbulence being responsible for the sub-structure that we see in {the type II burst fine structure}. For this we search for a power-law distributions of the size-scales associated with the inhomogeneity in the radio burst, similar to the analysis performed by \citet{chen2018, reid2021} for type III bursts.


\begin{figure*}[!ht]
    \begin{center}
\includegraphics[scale=0.33, trim=6cm 1cm 0cm 1cm]{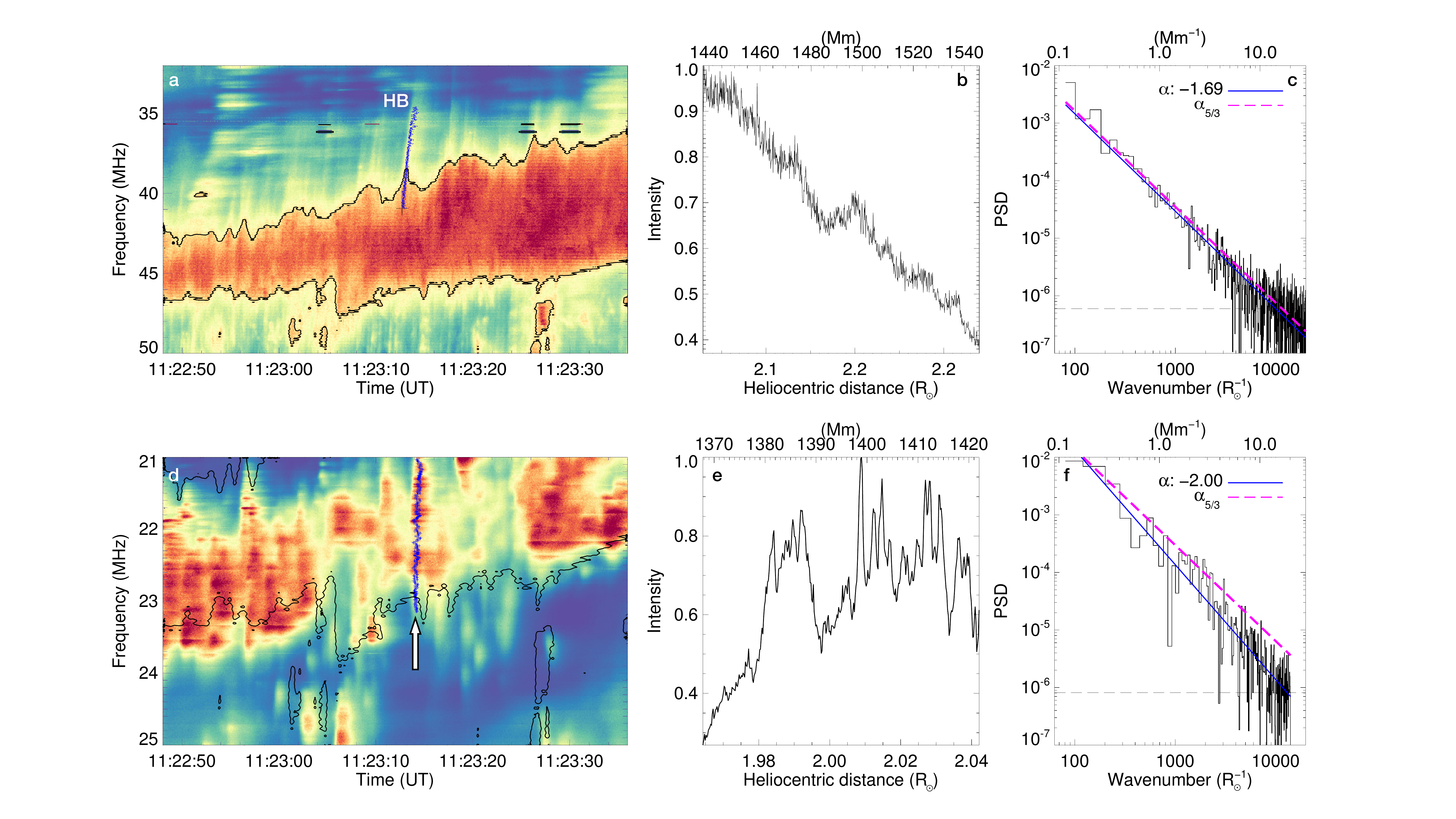}
\caption{PSD analysis of the herringbone radio burst for both its fundamental and harmonic component. (a) Herringbone radio bursts of {type II part-a}, with one herringbone market as `HB'. The contour is at 50\% of maximum intensity, chosen as an approximate outline of the backbone.  (b) Intensity as a function of distance for the `HB' herringbone. (c) PSD analysis of the intensity vs. distance profile, showing agreement with the Kolmogorov spectral index of $-5/3$. (d) {Fundamental component of the type II. The black contour from the harmonic (in panel a) is over-plotted on the fundamental. A drifting feature is marked by blue dots, starting at the white arrow.} (e) Intensity vs. distance for the drifting feature. (f) PSD analysis for the drifting feature, with a slightly steeper spectrum than the $-5/3$ value. }
\label{fig:psd_typeIIab}
\end{center}
\end{figure*}

Type II radio bursts are assumed to be plasma emission from mildly relativistic electrons accelerated at coronal shocks \citep{mann2005}. For plasma emission, the frequency of emission $f_{pe}$ is directly related to the electron number density $n_e$ in the corona ($f_{pe}\approx 8980\sqrt{n_e}$, where $n_e$ is in cm$^{-3}$ and $f_{pe}$ is in Hz). If we assume a coronal density that follows a hydrostatic equilibrium we may obtain an estimate for the altitude of the emission in the corona for any frequency using a density model. At any one time, the extent in frequency space of any spectral feature can also give the extent of the emission source in real space, provided we assume the density inhomogeneity is an enhancement of the background density model. {Previous numerical modelling has also shown that density turbulence modulates the level of Langmuir waves in the plasma emission process, e.g. \citet{reid2017, reid2021}}.
Hence we can perform a power spectral density (PSD) analysis of a radio burst intensity variation to obtain the distribution of size-scales of density perturbations. 


\subsection{Type II parts a and b: harmonic and fundamental}

Figure~\ref{fig:psd_typeIIab} shows our analyses for {parts-a and -b of the type II}. {Type II part-a} {is composed mostly of a series of} fine structures known as herringbone bursts. Herringbones are relatively rare, with only 20\% of type II bursts having this kind of fine structure \citep{cairns1987}. They consist of a series of forward and reverse fast-drifting bursts and are considered to be bursty electron acceleration at a coronal shock front, {with electron beams propagating in opposite directions away from the shock}. {The {type II part-a} emission lane} starts at $\sim$45\,MHz and drifts at a rate of $\sim$-0.1\,MHz\,s$^{-1}$, meaning the shock responsible for the herringbones had a start heliocentric distance of $\sim$2\,R$_{\odot}$ and speed of 1166\,km\,s$^{-1}$ using a Newkirk model \citep{newkirk1961} -- {this is close to the deprojected EUV wave speed of 950$\pm$310\,km\,s$^{-1}$}. 

{Type II part-b} has a different appearance, beginning at approximately 23\,MHz, with a more fragmented structure, see Figure~\ref{fig:psd_typeIIab}d. The frequency ratio of type II {parts-a and -b} means they {are likely the fundamental (F) and harmonic (H) pairs of the type II}. 
For example, the contour demarcating the backbone in Figure~\ref{fig:psd_typeIIab}a is overplotted on Figure~\ref{fig:psd_typeIIab}d. This shows {type II part-b} is the fundamental backbone component of the radio burst. Hence, a PSD analysis on the drifting components of {this F-H pair} provides an opportunity to determine turbulence characteristics close to the shock surface (observed from the fundamental backbone) and the upstream region into which electron beams propagate (observed from the herringbones).





\subsection{Type II parts a and b: power spectral density}
To perform the PSD, we extract an intensity vs. frequency profile from a prominent herringbone radio burst occurring at $\sim$11:23:11\,UT and starting at $\sim$43\,MHz, see blue points in Figure~\ref{fig:psd_typeIIab}a. 
As a pre-processing step, each spectrum is `flattened' by division of an empirical bandpass correction to account for the spectral response of NenuFAR, ensuring that any intensity enhancement is due to received flux rather than the variability in the system response across frequency. We then convert the frequency range to density using the Newkirk model, which provides us with an intensity vs. distance profile. The intensity profile was resampled by interpolating to an even distance grid with $\Delta x=0.07$\,Mm (distance equivalent of $\Delta f=6$\,kHz at 40\,MHz), see Figure~\ref{fig:psd_typeIIab}b. A PSD was performed on the intensity profile in order to obtain the distribution of coronal size-scales responsible for the herringbone burst, see panel c. A power law of the form $P(k)\sim k^{\alpha}$ is fit to the PSD, where $k$ is the wavenumber in units of inverse solar radii ($k=2\pi/\lambda_{R\odot}$) and $\alpha$ is the spectral index. A power law distribution of the size-scales of intensity (density) fluctuations is indicative of the scale invariance expected of a turbulent system. We find a spectral index of $\alpha=-1.69$, which matches the expectations of fully developed turbulence e.g., with $\alpha=-5/3=-1.67$ \citep{kolmogorov1941}. We performed the same PSD analysis for ten prominent herringbones, which resulted in an average spectral index of $\alpha_{\mu}=-1.71$. An average electron beam speed of 0.19\,c was deduced from the drift rate of the ten herringbones (using the Newkirk model), matching previous observations \citep{mann2005}. No significant difference was found for the spectral indices or speeds between forward and reverse drift herringbones. This shows that the shock responsible for these herringbone bursts accelerated electron beams into the corona, which then propagated through a turbulent medium as they induced plasma emission.



{The above result was tested with other density models that are commonly used for the metric wavelength range. For example, previous authors have highlighted the use for a 3.5$\times$Saito density model \citep{saito1977} for type II observations at metric wavelengths \citep{magdalenic2010, magdalenic2012, jebaraj2020}. Such a model also produces a power law distribution with an index of $\alpha=-1.72$, similar to the Newkirk model. Hence the result is not sensitive to the choice of density model.}

{Continuing with the Newkirk model, we then carry out the same analysis for a drifting component of the fundamental {type II part-b}, shown in Figure~\ref{fig:psd_typeIIab}d starting at 11:23:14\,UT and drifting from $\sim$23-21\,MHz. This drifting burst is not necessarily related to the herringbone in panel a. It is an independent measure of emission fine structure generated in the fundamental component close to the same time as the herringbone. The power spectrum of intensity vs distance for this drifting structure again shows a powerlaw distribution but with a steeper index of $\alpha=-2.0$}. 

While the herringbone of the harmonic represents a beam propagation into the unshocked upstream corona, the fundamental backbone emission is likely a sample of the density turbulence closer to the shock surface, and its steeper PSD index may be an indicator of slightly different turbulence characteristics closer to the shock front; we discuss this further in Section 4.3.

As for the size-scales of the turbulence, we find the power law exists over a range of wave numbers from $\sim$0.1 to 30\,Mm$^{-1}$, similar to the wavenumbers reported in \citet{chen2018}. This means the distribution of size-scales for the density inhomogeneities varies from 62\,Mm to as small as 209\,km. Values of the outer scale of density turbulence in the corona are found to be on the order of 696\,Mm (1\,R$_{\odot}$) at a radial distance of $\sim$7\,R$_{\odot}$ using observations from the \emph{Ulysses} and \emph{Galileo} missions \citep{wohlmuth2001}, while the inner scales of turbulent energy dissipation at $\sim$2\,R$_{\odot}$ are expected to be $<$1\,km  \citep{coles1989, sasiraja2019}. This means the size-scales we derive here are in the inertial range, between the inner and outer scale of turbulence. In Section 4.4 we discuss the potential of NenuFAR to provide observation close to the inner-scale (on the order of kilometers), where energy dissipation is expected to occur. 

{Finally, we may estimate the amplitude of the density perturbations from the intensity perturbations in the fundamental component of the radio burst (type II part-a), using the expression derived in \citet{reid2021}
\begin{equation}
\frac{\Delta n}{n} = \frac{v_{th}^2}{v_b^2}\frac{\Delta I}{I}
\end{equation}
where $\Delta n/n$ and $\Delta I/I$ are the fractional density and intensity perturbations, respectively, $v_{th}$ is the thermal speed of the plasma and $v_b$ is the electron beam speed. We assume a 2\,MK plasma, resulting in $v_{th}=5.5$\,Mm\,s$^{-1}$, and the exciter speed of the drifting feature in type II part-a is found to be 78\,Mm\,s$^{-1}$ using a Newkirk density model. 
The intensity perturbations across the drifting feature produce $\Delta I/I\sim0.14$, providing a $\Delta n/n\sim 0.8\times10^{-3}$, which {is somewhat similar} to the value of $3\times10^{-3}$ found from type III striae in \citet{reid2021}. 
{At larger heights than those observed here}, previous radio scintillation observations have shown similar values $\Delta n/n=10^{-3}$ at $\sim$10\,R$_{\odot}$, with the fluctuations increasing to between $10^{-2} -10^{-1}$ at distances out to 0.8\,AU \citep{woo1995, sasi2016}.}

\begin{figure}[t!]
\includegraphics[scale=0.4, trim=2cm 7cm 0cm 0cm]{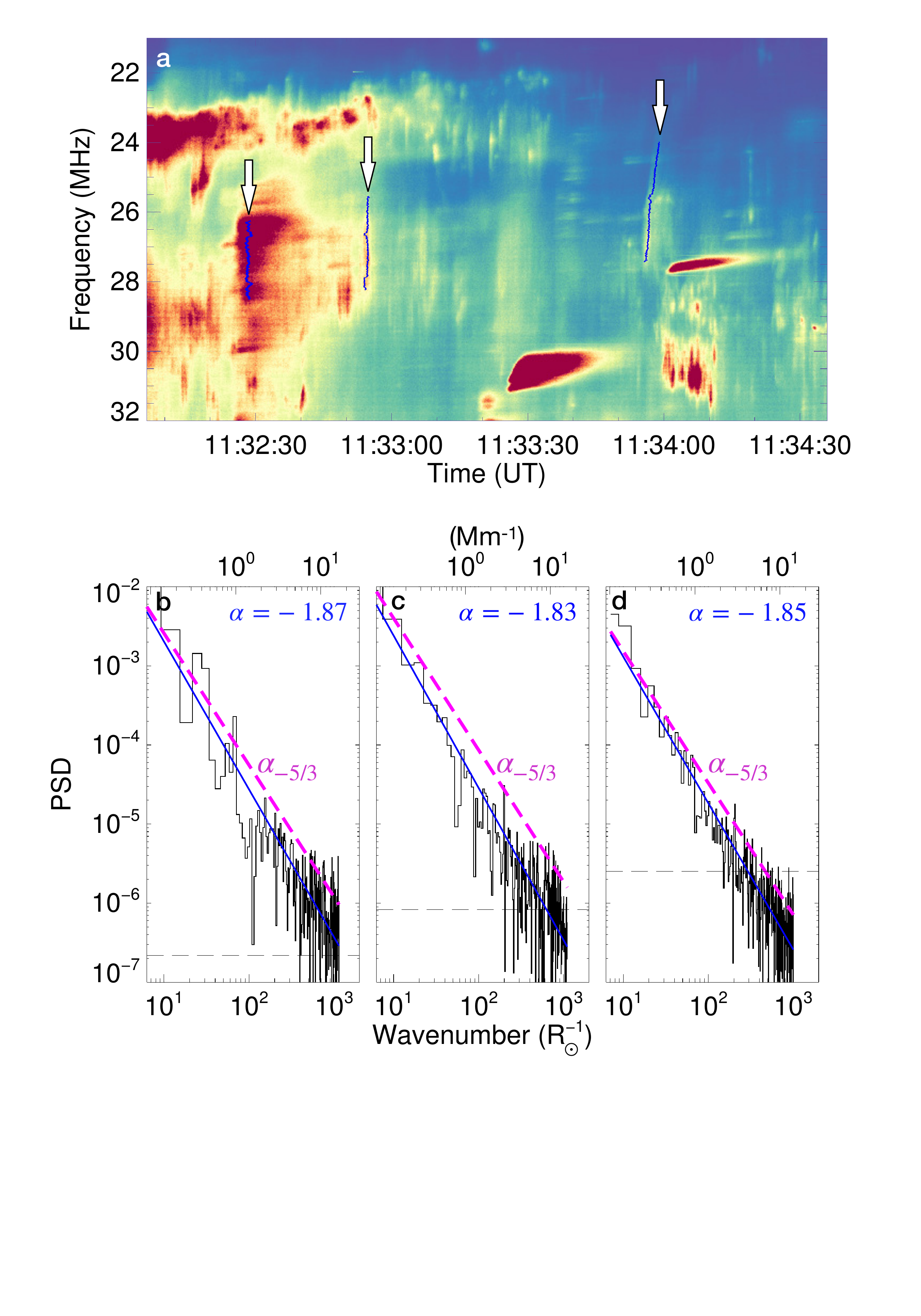}
\caption{{(a) Zoom of the region showing the {type II part-c} part of the burst (harmonic component), displaying fine structure and fragmentation. We have identified three drifting features among this fine structure, indicated by the arrows. An intensity vs frequency profile is extracted along these bursts (at the indicated blue points), used to perform the same PSD analysis as above. (b-d) A PSD for each of the drifting features in the dynamic spectrum. They again show a powerlaw distribution, but with slightly steeper spectral indices ($\alpha=-1.83$ to $-1.87$) than for the harmonic herringbones.}}
\label{fig:psd-typeIIc}
\end{figure}




%
%
\vspace{1em}
\subsection{{Type II part-c}: Power spectral density}

{Part-c of the type II burst} lasts from 11:28-11:38\,UT starting at a frequency of $\sim$30\,MHz and drifting at a rate of -0.022\,MHz\,s$^{-1}$, see Figure~\ref{summary-fig}. Given its position in the dynamic spectrum, {type II part-c} is  a continuation of {part-a of the radio burst}, and we consider it to be a harmonic component. Using the Newkirk density model, the start frequency of 15\,MHz and drift rate give a shock heliocentric distance and speed of 2.6\,R$_{\odot}$ and 538\,km\,s$^{-1}$, respectively.
Hence this part of the type II represents the shock at a larger altitude than the herringbone burst and at a slower speed.

{Unlike the herringbone burst, type IIc is more fragmented and has fewer discernible drifting features, but we were able to identify three drifting bursts, see Figure~\ref{fig:psd-typeIIc}a. Performing the same power spectrum analysis as above  shows a powerlaw distribution of intensity perturbations, see Figure~\ref{fig:psd-typeIIc}b-d. The powerlaw in each is slightly steeper than the herringbone harmonic burst, showing values of $\alpha=1.83-1.87$. The steeper index may indicate that the radio emission comes from a different region in the corona with different turbulence characteristics to the region that produced the initial herringbones; this is discussed further in Section 4.3.}

From Figure~\ref{summary-fig}, {type II part-c} does not appear isolated but is embedded in the low frequency end of a broad-band feature consisting of a number of faint forward and reverse drift bursts occurring at 30--55\,MHz and lasting from $\sim$11:28--12:40\,UT. The frequency range is indicative of an heliocentric distance of 1.5--1.8\,R$_{\odot}$ using a Newkirk model {(assuming this is fundamental emission)}, and the burst frequency drifts give an exciter speeds of 0.05\,c. A PSD analysis gives $\alpha=-1.64$ for the drifting bursts here, which is again a signature of turbulent plasma. {This broad-band feature is likely a part of  the type IV burst that can also be observed in the ORFEES} dynamic spectrum from 11:20 to 11:45\,UT, extending up to frequencies of $\sim$600\,MHz. The type IV is indicative of energetic electrons trapped in flare loops or associated with a CME (c.f \citet{carley2016, carley2017, morosan2019}.  {While the type IV burst} may be from plasma emission in a turbulent environment, the origin of the emission and associated electron acceleration is unclear due to the lack of images of the radio source. In future studies, the inclusion of imaging along with a PSD diagnostic of type IV fine structure could be particularly useful for analysing the turbulent plasma properties of flare loops and CMEs.



%
%

\section{Discussion}

 \begin{figure}[t!]
    \begin{center}
\includegraphics[scale=0.38, trim=2cm 8cm 0cm 1cm]{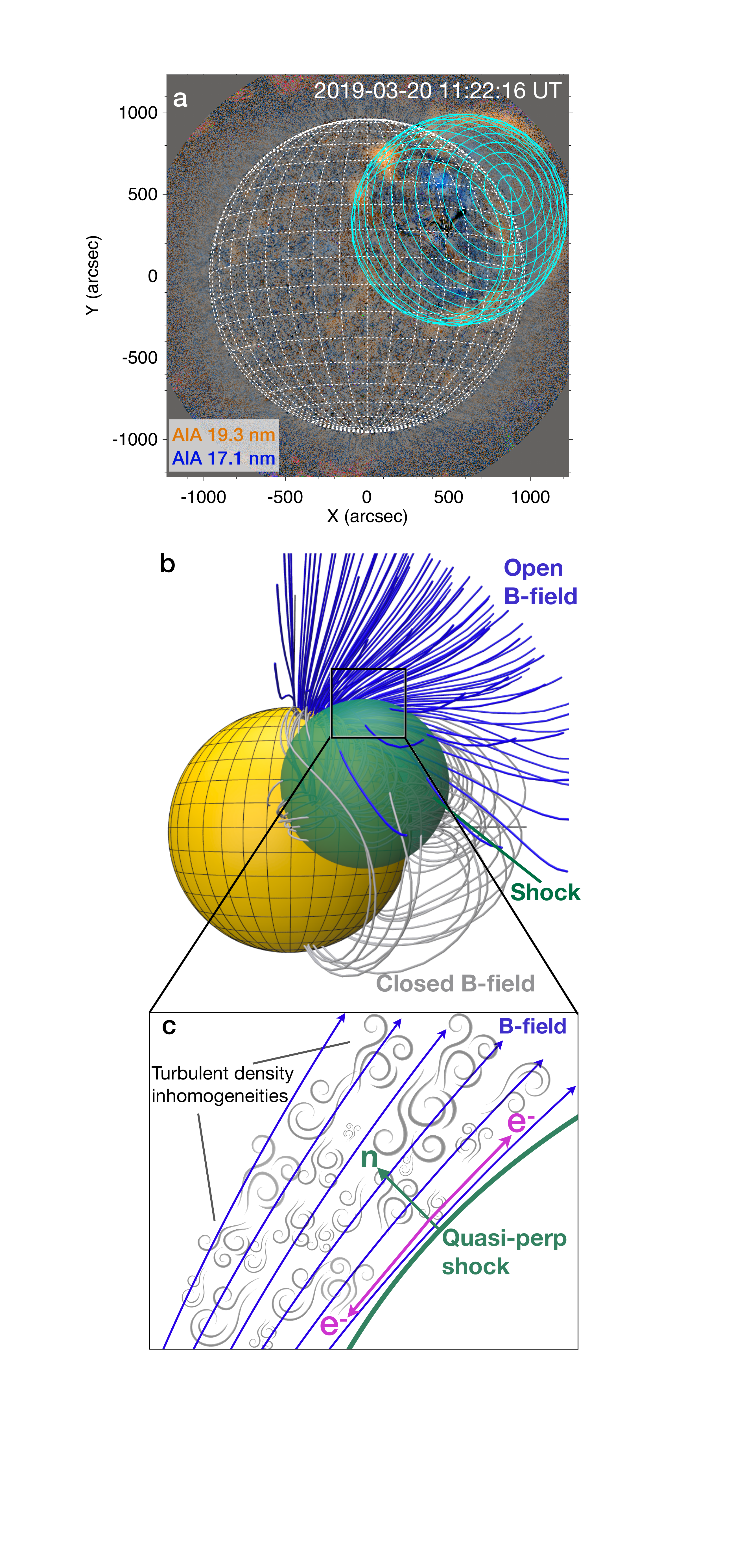}
\caption{ {(a) The EUV wave propagating on-disk and off-limb is approximated by a 3D spheroid in the solar atmosphere, projected onto the AIA image (turquoise lines). (b) This 3D spheroid is shown as the shock surface {(green sphere) embedded in a PFSS extrapolation of the magnetic field. There was a large number of open field lines to the north of the active region, where the shock normal to the magnetic field is quasiperpendicular}.} (c) An illustration of the turbulent environment into which electron beams are accelerated in opposite directions.}
\label{fig:model}
\end{center}
\end{figure}

\subsection{Shock location and geometry}

The herringbone analysis provided above shows that shocks in the corona can produce bursty acceleration of electrons in a turbulent coronal environment. Unfortunately no radio imaging observations were available at the time of these radio bursts, so we cannot say where in the corona these electron beams and turbulence were located. However, type II bursts and herringbones are expected to occur in a shock with quasiperpendicular ($\sim\perp$) geometry at the flanks of an eruption \citep{carley2013, morosan2019}. Given that we can image the shock propagation in EUV, we have the opportunity to determine where this disturbance may have encountered $\sim\perp$ shock geometry.

{In Figure~\ref{summary-fig}a-b a disturbance {is visible} in AIA, propagating both on disk and and off-limb in the shape of a dome-like structure. We assume this disturbance to be a signature of the shock in EUV images. To estimate the extent of this shock, we reconstruct a spheroid in 3D space and project it onto the AIA image at the start time of the type II in NenuFAR ($\sim$12:22:16\,UT). The shape and position of the spheroid is adjusted by-eye such that its extent across the solar surface and off-limb matches the regions of EUV emission in the AIA images, see Figure~\ref{fig:model}a. The EUV disturbance is well described by this spheroid and allows us to roughly demarcate the 3D extent of the `shock bubble'. At $\sim$12:22:16\,UT the bubble apex was at a heliocentric distance of $\sim$2.1\,R$_{\odot}$.}

{Figure~\ref{fig:model}b shows this shock bubble embedded in a potential field source surface (PFSS) extrapolation of the coronal magnetic field \citep{Stansby2020} using data from the Global Oscillation Network Group \citep[GONG;][]{Harvey1996}.} It shows that a significant amount of open field existed at the flanks of the eruption towards solar north, with the orientation of this field with respect to the shock being $\sim\perp$. While we cannot directly image the herringbone bursts, we assume the region to the north of the shock to be the most probable place for electron acceleration. Figure~\ref{fig:model}c shows an illustration of how a herringbone burst with a turbulent signature may be generated e.g., with bi-directional electron beams being accelerated into turbulent plasma on open field in the upstream region of the shock. 

\subsection{Electron beam acceleration in coronal turbulence}

There are a variety of particle acceleration mechanisms that have been proposed to explain the presence of high energy particles produced by plasma shocks. One of the most common is through the so-called `first order Fermi mechanism' in which a charged particle undergoes repeated reflections between the upstream-downstream environment, gaining energy upon each transition \citep{axford1977, drury1983}. This diffusive shock acceleration mechanism is usually employed for protons or ions, as their large gyro-radii mean they can interact with turbulent magnetic fluctuations either side of the shock to produce the repeated reflections. The particle energy gain can reach GeV energies \citep{vainio2009}, while the mechanism efficiency can depend on the shock geometry \citep{jokipii1987}.

For electrons the acceleration mechanism remains less clear. At low (thermal) energies electron gyro-radii are too small for interaction with magnetic perturbations in the background plasma, so they cannot experience the repeated reflections necessary for the diffusive Fermi mechanism. In order to explain how electrons are accelerated to non-thermal energies by a shock, the shock-drift acceleration (SDA) mechanism is employed, in which the electron gains energy while undergoing a magnetic mirroring combined with a $\nabla B$-drift in the presence of the convective $E=\vec{v}\times \vec{B}$ electric field of the shock \citep{wu1984, ball2001}. This mechanism has been used to explain type II radio bursts as it easily produces electrons of moderate non-thermal energies \citep{holmon1983}. 

With the addition of turbulence in SDA, the electrons may encounter the shock multiple times and gain higher energies than just a single shock reflection \citep{burgess2006}. In the herringbones observed here, analysis of the electron beam energies can provide clues as to whether turbulence is involved in the acceleration mechanism. 
{As stated above, the drift of the herringbones in Figure~\ref{fig:psd_typeIIab}a provide} electron beam speeds of $v_{beam}=0.19$\,c (using a Newkirk model), {meaning the maximum speed in the electron beam distribution is $v_{max}=2v_{beam}=0.38$\,c, or 41\,keV. According to \citet{ball2001} the maximum energy gain of a particle reflected from a shock through the SDA mechanism is $E_r/E_i$$\leq$13.93, where $E_r$ and $E_i$ are the particle reflected and incident energy, respectively. If we take the incident energy to be a few times the thermal kinetic energy of a 1\,MK plasma, then $E_i$$\sim$1\,keV, meaning the electrons producing {type II part-a} would need to experience an energy gain of up to $E_r/E_i$$=$41. This is beyond the single reflection limit, and would require multiple reflections of the SDA process. Several authors have modelled such multiple reflections from ripples on the shock surface in response to turbulence in the background plasma \citep{burgess2006, guo2010}. In our case, the observed turbulence during herringbone production could be responsible for electron beam energy gain to 41\,keV.}

A statistical comparison of turbulence characteristics (amplitude and spectral index) to herringbone kinematics would help in confirming any relationship between the two phenomena. Unfortunately only ten herringbones in part-a of the type II were clear enough to obtain this information, so we cannot perform such a statistical study at present. We note, however, that the PSD spectral index of the herringbone features is different to the features of {type II part-c}, which shows far fewer drifting features. Particular kinds of shock inhomogeneity and turbulence in the corona may be responsible for the production of herringbones. Further studies of these bursts in the context of turbulence are required, especially those which include turbulence diagnostics through imaging observations \citep{subra2011}.

\subsection{Evidence for different kinds of turbulent coronal environments?}

The herringbone radio bursts show a turbulence signature of the classical Kolmogorov type, with a spectral index of $\alpha=-1.71$. However drifting features in {part-c of the type II} have a steeper index of $\alpha=-1.85$ and {type II part-a} is even steeper at $\alpha=-2.0$. What is the cause of these different spectral indices? 
While a steeper spectral index might be encountered beyond the turbulence inner-scale, this is expected to occur at $<$1\,km at a heliocentric distance of 2\,R$_{\odot}$ \citep{sasiraja2019}. Given we observe size-scales $>$209\,km, we may rule out the presence of steep indices due to observation beyond the inner-scale and discuss indices values usually observed for inertial scales.

The turbulence spectral index for inertial scales depends on a variety of factors, including the turbulent property under investigation e.g., whether it is velocity, magnetic field, or density perturbations. The original Kolmogorov formulation of $k^{-5/3}$ is predicted for velocity perturbations in an incompressible flow in a neutral medium \citep{kolmogorov1941}. An extension of this formulation to the compressible magnetohydrodynamic case predicts a variety of possible spectral indices which may range from $\alpha=-1$ to $-2$ \citep{matthaeus1982, yamauchi1998}. 
This range of spectral indices is reflective of the various MHD wave modes a plasma may support e.g., Alfv\'{e}n waves, fast and slow MHD waves. These wave modes lead to compressibility and anisotropy of the turbulence and can result in a departure from the incompressible, isotropic and neutral Kolmogorov value of $k^{-5/3}$ \citep{saur2002, shaikh2010, kowal2007}. 

For density perturbations, which we observe here through radio burst fine structure, the story is more complex. In the `nearly incompressible' case dominated by magnetic fields, the density perturbations should theoretically scale similarly to pressure and follow a $k^{-7/3}$ (which assumes a polytropic equation of state $p\sim\rho^{\gamma}$, where $\gamma$ is the adiabatic index). However, observational work has shown that density turbulence in the corona and solar wind often follow the Kolmogorov value of $k^{-5/3}$ from $2-40$\,R$_{\odot}$ \citep{scott1983, coles1989, armstrong1990}. That said, MHD simulations from \citet{kowal2007} have shown super-Alfvénic flow will produce a density spectrum of $k^{-7/3}$, while low Mach numbers result in $k^{-5/3}$. The presence of strong magnetic fields also introduces an anisotropy into the density spectra, with perturbations perpendicular to the background field following  $k_{\perp}^{-2}$, while those parallel to the field follow $k_{||}^{-5/3}$ \citep{kowal2007}. {Furthermore, \citet{reid2021} have recently shown that finite Langmuir wave group velocity during the plasma emission process can smooth-out finer-scale variability in the emission at high spatial wavenumbers, resulting in a steepening of the spectral index in the inertial range to values less than $-5/3$.}

{It's clear the observed spectral index for density turbulence can take on a variety of values depending on the specific conditions of the plasma environment. This may explain the different morphologies in each part of the type II} we observe in the event reported here. {For example, the herrinbones are from electron beam propagation some distance away from the shock \citep{carley2015, morosan2019}, and likely provide a measure of density turbulence in the background corona. The drifting feature in Figure~\ref{fig:psd_typeIIab}d crosses the fundamental emission lane, indicating the exciter may be closer to the shock surface; this region should be a more perturbed environment and larger density compressions would lead to a steeper turbulence index \citep{kowal2007}. At the later time of {type II part-c} the shock may have reached a different region of the corona (it occurred at a larger distance of 2.6\,R$_{\odot}$), having its own characteristic turbulence signature. This is perhaps expected, given that different regions of the solar atmosphere have been shown to have the different turbulent spectral indices ranging from $k^{-1.7} - k^{-2.3}$ \citep{abramenko2005}}. 
Radio sounding experiments have also shown a steepening of density power spectra near the Kolmogorov value during the passage of a CME in the heliosphere from $<$10\,R$_{\odot}$ to 50\,R$_{\odot}$ \citep{woo1992, efimov2008}.


Finally, the different appearance of {type II part-c} may also be due to the magnetic environment in which the shock is generated. For example, {parts-a and -b of the radio burst} were likely from {a quasi-perpendicular shock-geometry} as described above, which produces efficient electron acceleration.  On the other hand, {type II part-c} occurred at a higher altitude where there {are less quasi-perpendicular field orientations} and hence less opportunities for electron acceleration; this may explain why {type II part-c} is {a weaker section of the} radio burst, similar to the analysis of \citet{maguire2020}.


\subsection{Spatial scales of turbulence}

As for the observed spatial scales in the inertial range of turbulence, these are limited by the spectral resolution of the instrument and the signal-to-noise of fine scale spectral features. We observe spatial scales from 209\,km to 62\,Mm, in agreement with previous observations from LOFAR \citep{chen2018}. The smallest scale that we observe is at least two orders of magnitude greater that the expected inner scale of 0.5-1\,km over which turbulent energy dissipation occurs in the low corona \citep{coles1989}. At this scale, the spectral index of the PSD would be expected to decrease from the inertial range ($-5/3$) to $<$$-2$, which is associated with the dissipative energy range of turbulence. 

The questions is, can NenuFAR or LOFAR offer enough frequency resolution to observe the scales of turbulent energy dissipation in the corona using the technique outlined here?
At 85\,MHz (the upper boundary of NenuFAR's observing range), a 6\,kHz spectral resolution results in a spatial domain resolution of 17\,km in the corona, while the same spectral resolution at 300\,MHz results in a spatial resolution of 3.8\,km. Hence, the characteristics of instruments such as NenuFAR and LOFAR operate on the margins of being able to resolve the spectral steepening indicative of the energy dissipation associated with coronal turbulence. This of course assumes that plasma emission can be observed at such fine frequency resolution.  
Future studies should consider such experimentation, as observation of plasma emission and density inhomogeneity could provide insight into turbulent energy dissipation on spatial scales at which kinetic energy is thermalized and the corona is heated, see e.g. simulations by \citet{sokolov2013}.

\section{Conclusion}

In this paper we used a PSD analysis of NenuFAR data to diagnose the distribution of spatial size-scales of density turbulence in the corona that were responsible for different {types of fine structure in a type II radio burst}. The initial harmonic component ({type II part-a}) showed the most promising evidence for turbulent structure. Its herringbone fine structure showed a powerlaw PSD of spectral index $\alpha=-1.71$, which is close to the value of $-5/3$ expected for fully developed turbulence. It is likely that this turbulence existed in the coronal background, with the shock wave passing through it.

The other {parts of the type II burst} showed a powerlaw PSD signature, but with indices steeper than than the Kolmogorov value. This may be indicative of the different levels of density perturbations at the shock-front or a propagation in different coronal environments.

The high time resolution, spectral resolution (and by proxy spatial resolution) that new radio telescopes such as NenuFAR and LOFAR are able to observe offer for the first time a means of determining the nature of turbulence in the corona at a range of heights and in a host of different environments i.e., shocks, flares, and CMEs, among others. More work is needed to determine the proportion of type II radio bursts (and other types) that show this turbulent signature, allowing us to diagnose particle acceleration in the turbulent environment of coronal shocks.

\section*{Acknowledgments}
EPC is supported by the Schr\"odinger Fellowship at DIAS. KSR acknowledges the financial support from the Centre National d'\'{e}tudes Spatiales (CNES), France. This paper is based on data obtained using the NenuFAR radiotelescope. NenuFAR has benefited from the following funding sources: CNRS--INSU, Observatoire de Paris--PSL, Station de Radioastronomie de
Nançay, Observatoire des Sciences de l'Univers de la Région Centre, Région Centre-Val de Loire, Université d'Orléans, DIM-ACAV and DIM-ACAV+ de la Région Ile de France, Agence Nationale de la Recherche. We acknowledge the use of the Nançay Data Center (CDN), hosted by the Nançay Radio Observatory (Observatoire de Paris--PSL, CNRS, Université d’Orléans), and also supported by Région Centre-Val de Loire. Support from Paris Astronomical Data Centre (PADC) is acknowledged for data storage and from LESIA for computing capabilities. We would like to thank GOES and SDO teams for data access. Finally, we would like to thank the referee for their useful comments and suggestions during the review of this manuscript.

\bibliography{nfar_typeII.bib}{}



\end{document}